\documentclass[english,12pt]{article}
\usepackage{color}
\usepackage{amssymb}
\usepackage{amsmath}
\usepackage{babel}
\usepackage{pifont}
\usepackage
[dvips,
colorlinks=true,
linkcolor=webgreen,
filecolor=webbrown,
citecolor=webgreen,
bookmarksopen=false,
]{hyperref}
\definecolor{webgreen}{rgb}{0,.5,0}
\definecolor{webbrown}{rgb}{.6,0,0}
\usepackage[T1]{fontenc}
\usepackage[cp1250]{inputenc}
\usepackage{array}
\usepackage[dvips]{graphicx}
\usepackage{amssymb}
\date{}
\definecolor{arcolor}{cmyk}{0.05,0.95,0.9,0.1}
\title{Quantum solution to the Newcomb's paradox}
\author{Edward W. Piotrowski\\ Institute of Theoretical Physics,
University of Bia\l ystok,\\ Lipowa 41, Pl 15424 Bia\l ystok,
Poland\\ e-mail: \href{mailto:ep@alpha.uwb.edu.pl}{ep@alpha.uwb.edu.pl}\\
 Jan S\l adkowski\\ Institute of Physics, University of Silesia, \\ Uniwersytecka
4, Pl 40007 Katowice, Poland \\ e-mail:
\href{mailto:sladk@us.edu.pl}{sladk@us.edu.pl} }
\begin{document}
\maketitle
\def\Z{{\bf Z\!\!Z}}
\def\R{{\bf I\!R}}
\def\N{{\bf I\!N}}
\begin{abstract}
We show that quantum game theory offers solution to the famous
Newcomb's paradox (free will problem). Divine foreknowledge is not
necessary for successful completion of the game because quantum
theory offers a way to discern human intentions in such way that
the human retain her/his free will but cannot profit from changing
decision. Possible interpretation in terms of quantum market games
is proposed.

\end{abstract}

 \vspace{5mm}

\section{Introduction}

There is a common belief that the characteristic size of the
brain's integral parts is too big to allow for quantum effects
being important \cite{1}. But recent experiments show that
separated objects of the size of a golf ball can form quantum
entangled states even in a room temperature \cite{2}. Physicists
successfully apply quantum mechanics to describe a lot of complex
system that may have in principle arbitrary size, including black
holes or even the whole Universe. Are there any reasons for
quantum modeling of phenomena related to brain activity,
consciousness or social behaviour? One can give an answer to this
question only after construction and thorough verification of
respective models \cite{3}. Below we consider a problem easily
susceptible of modeling as a quantum game that should shed
some light on the solutions that quantum theory may offer. \\

In 1960 William Newcomb, a physicist, intrigued  the philosopher
Robert Nozick with a claim that in an elementary game
characterized by the matrix $M$
\begin{equation}
M:=\begin{pmatrix}\$ 1000& \$ 1\,001\,000\\
\phantom{\$ 100}0&\$ 1\,000\,000
\end{pmatrix}
\label{machnewcomb}
\end{equation}
giving the pay-off of the player 1 in all possible situations, the
player 1 is not able to chose his strategy without having any
measure of occurring {\it a posteriori} of any of the four
possible events. Rows correspond the player 1's strategies:
feminine $|\mathsf{0}\rangle_1$ and masculine
$|\mathsf{1}\rangle_1$\footnote{The use of the adjectives feminine
and masculine to underline the character of the strategies will be
explained later, see also the Gardner book} and columns to
opponent's strategies $|\mathsf{0}\rangle_2,
|\mathsf{1}\rangle_2$. It so happens even despite the fact that
the feminine strategy dominates the masculine one (that is the
pay-off is greater regardless of the opponents strategy). The
choice of the masculine strategy $|\mathsf{1}\rangle_1$ is more
profitable when the event corresponding to the off-diagonal
elements of $M$ do not occur and the rest have almost equal
probabilities. This might happen if the opponent is able to
foresee the player 1 moves. Due to this paradoxical property the
above game with indefinite (hidden) set of occurrences became for
philosophers, economists and theologians a graceful theme of
speculations about free will and its consequences \cite{4,5}. The
disputes, often referred to as newcombmania \cite{6}, deserve a
thorough analyzis from the quantum game theory point of view
\cite{7}-\cite{11}. The development of the probability theory
provide us with many intriguing examples where ambiguous
specification of the appropriate probability measures resulted in
contradiction (Bertrand \cite{13} and Banach-Tarski \cite{14}
paradoxes are the most famous ones). One can still find people who
regardless of this facts continue philosophical disputes while
ignoring the necessity of precise definition of the probabilistic
measures in their models. We would like to show that quantum
theory may be of help in settling the ambiguities.
\section{Quantum description of the game}

Quantum game theory exploits the formalism of quantum mechanics in
order to offer the players new classes of strategies. Interesting
generalization of well known classical games have been put forward
\cite{7,8}. There are arguments that quantum strategies may offer
extraordinary tools for biologists \cite{15}-\cite{17}. Economics
being the theater of various games and conflicts should not
despise these new ideas \cite{18,12}. We will describe player's
strategies as vectors (often referred to as states) in Hilbert
spaces $\mathcal{H}_{i}$ where the subscripts $i=1,2$ distinguish
between the player 1 and 2. It is convenient to define the
strategy density operator $\mathcal{W}$
\begin{equation*}
\mathcal{W}=\sum_{r,s=1}^2W_{rs}
|\mathsf{r\negthinspace-\negthinspace1}\rangle_1\,|\mathsf{s\negthinspace-
\negthinspace1}\rangle_2\,{_1}\langle\mathsf{r\negthinspace-\negthinspace1}
|\,_2\langle\mathsf{s\negthinspace-\negthinspace1}|
\end{equation*}
where $(W_{rs})$ is a matrix with nonnegative entries such that
$\sum_{r,s} W_{rs}=1$ and
$|\mathsf{r}\rangle_1\,|\mathsf{s}\rangle_2\,{_1}\langle\mathsf{r}
|\,_2\langle\mathsf{s}| $,
$r,s\negthinspace\in\negthinspace\negthinspace\{\mathsf{0},\mathsf{1}\}$
are projective operators on the states of the game,
$|\mathsf{r}\rangle_1|\mathsf{s}\rangle_2\in\mathcal{H}_1\negthinspace
\otimes\mathcal{H}_2$. For our aims it will be sufficient to use
two dimensional Hilbert spaces for the players' strategies. The
states of the classical setting (mixed strategies) are represented
by a diagonal matrix $(W_{rs})$. Non-diagonal elements of
$(W_{rs})$ describe situations (strategies) that are out of the
reach for classical players. Following the classical terminology
we will call the pay-off observable $\mathcal{M}$ a Hermitian
operator corresponding to the matrix  $(\ref{machnewcomb})$:
\begin{equation*}
\mathcal{M}:=\sum_{r,s=1}^2M_{rs}
|\mathsf{r\negthinspace-\negthinspace1}\rangle_1\,|
\mathsf{s\negthinspace-
\negthinspace1}\rangle_2\,{_1}\langle\mathsf{r\negthinspace-\negthinspace1}
|\,_2\langle\mathsf{s\negthinspace-\negthinspace1}|\,.
\end{equation*}
Therefore, according to the classical interpretation of the game,
the player 1's expected pay-off is equal to the sum of diagonal
elements (trace) of the product of $M$ and the transpose of $W$:
$$
E(\mathcal{M}):=\text{Tr}\mathcal{MW}=\sum_{rs}M_{rs}W_{rs}=\text{Tr}
MW^T\,.
$$
\section{Newcomb's paradox}

M. Gardner proposed  the following fabulous description of a game
with pay-off given by the matrix $(\ref{machnewcomb})$ \cite{4}.
An alien Omega (or Alf?) being a omniscent representative of alien
civilization (player 2) offers a human (player 1) a choice between
two boxes. The player 1 can take the content of both boxes or only
the content of the second one. The first one is transparent and
contains \$1000. Omega declares to have put into the second box
that is not transparent \$1000000 (strategy
$|\mathsf{1}\rangle_2$) but only if he foresaw that the player 1
decided to take only the content of that box
($|\mathsf{1}\rangle_1$). A male player 1 thinks: {\it If Omega
knows what I am going to do then I have the choice between \$1000
and \$1000000. Therefore I take the \$1000000 }(strategy
$|\mathsf{1}\rangle_1$). A female player 1 thinks: {\it Its
obvious that I want to take the only the content of the second box
therefore Omega foresaw it and put the \$1000000 into the box. So
the one million dollar {\bf is} in the second box. Why should I
not take more -- I take the content of both boxes} (strategy
$|\mathsf{0}\rangle_1$). The question is whose strategy, male's or
female's, is better? One cannot give unambiguous answer to this
question without precise definition of the measures of the events
relevant for the pay-off.
\section{Human's and Omega's strategies}

Omega as representative of an advanced alien civilization is
certainly aware of quantum properties of the Universe that are
still obscure or mysterious to humans. The boxes containing
pay-offs are probably coupled. One can suspect this because the
human cannot take content of the transparent box only (\$1000).
The female player is sceptical about the possibility of
realization of the Omega's scenario for the game.  She thinks that
the choice of the male strategy results in Omega putting the one
million dollar in the second box, and after this being done no one
can prevent from her taking the content of the both boxes in
question (ie \$1001000). But Meyer proposed a quantum tactics
\cite{7} that, if adopted by Omega, allows  Omega to accomplish
his scenario. Let us note that Omega
 may not be able to foresee the future \cite{4}. For it
aims it is sufficient that it is able to discern human intentions
regardless of their will or feelings on the matter. The obstacles
to this implied by the no-cloning theorem can be overcome by means
of teleportation \cite{19}: Omega has must be able to intercept
and then return human's strategies. The presented below
manipulations leading to thwarting humans are feasible with
contemporary technologies. The course of the game may look as
follows. At the starting-point, the density operator $\mathcal{W}$
acting on
$\mathcal{H}_1\negthinspace\negthinspace\otimes\negthinspace\mathcal{H}_2$
describes the human's intended strategy and the Omega's strategy
based on its prediction of human's intentions. The actual game
must be carried on according to quantum rules that is players are
allowed to change the state of the game by unitary action on
$\mathcal{W}$ \cite{7,8}. The human player can only act on her/his
$q$-bit Hilbert space $\mathcal{H}_1$. Omega's tactics must not
depend on the actual move performed by the human player (it may
not be aware of the human strategy): its moves are performed by
automatic device that couples the boxes. The Meyer's recipe leads
to:
\begin{enumerate}
\item Just before the human's move, Omega set the automatic devise
according to its knowledge of human's intention. The device
executes the tactics $\mathcal{F}\negthinspace\otimes\mathcal{I}$,
where $\mathcal{I}$ is the identity transform (Omega cannot change
its decision) and $\mathcal{F}$ is the well known Hadamard
transform frequently used in quantum algorithms:
$F:=\tfrac{1}{\sqrt{2}}\begin{pmatrix}
1&\phantom{-}1\\1&-1\end{pmatrix}$.
\item The human player with the probability  $w$ uses the female
tactics $\mathcal{N}\negthinspace\otimes\mathcal{I}$\,, where
$\mathcal{N}$ is the negation
operator\footnote{$\mathcal{N}|\mathsf{0}\rangle=|\mathsf{1}\rangle$,
$\mathcal{N}|\mathsf{1}\rangle=|\mathsf{0}\rangle$} and with the
probability $1\negthinspace-\negthinspace w$ the male tactics
$\mathcal{I}\otimes\mathcal{I}$.
\item At the final step the boxes are being opened and the built-in
coupling mechanism performs once more the transform
$\mathcal{F}\otimes\mathcal{I}$ and the game is settled.
\end{enumerate}

\section{The course of the game and its result}
Let us analyze the evolution of the density operator
$\mathcal{W}$. The players' tactics, by definition, could have
resulted in changes in the (sub-)space $\mathcal{H}_1$ only
therefore it suffices to analyze the human's strategies. In a
general case the human can use a mixed strategy: the female one
with the probability $v$ and the male one with the probability
$1\negthinspace-\negthinspace v$. Let us begin  with the extreme
values of $v$ (pure strategies). If the human decided to use the
female strategy ($v\negthinspace=\negthinspace1$) or the male one
($v\negthinspace=\negthinspace0$) then the matrices
$\mathcal{W}_i$, $i=0,1$ corresponding to the density operators
\begin{equation*}
\mathcal{W}_0=\sum_{r,s=1}^2{W_0}_{rs}
|\mathsf{r\negthinspace-\negthinspace1}\rangle_1\,|\mathsf{0
}\rangle_2\,{_1}\langle\mathsf{s\negthinspace-\negthinspace1}
|\,_2\langle\mathsf{0}|\,
\end{equation*}
and
\begin{equation*}
\mathcal{W}_1=\sum_{r,s=1}^2{W_1}_{rs}
|\mathsf{r\negthinspace-\negthinspace1}\rangle_1\,|\mathsf{1
}\rangle_2\,{_1}\langle\mathsf{s\negthinspace-\negthinspace1}
|\,_2\langle\mathsf{1}|\,
\end{equation*}
are calculated as follows: \vspace{1ex}
\begin{equation*}
\begin{split}
&\hspace{2em}\begin{pmatrix}
v&0\\
0&1-v
\end{pmatrix}
\longrightarrow  \tfrac{1}{2}
\begin{pmatrix}
1&\phantom{-}1\\1&-1
\end{pmatrix}
\negthinspace
\begin{pmatrix}
v&0\\
0&1\negthinspace-\negthinspace v
\end{pmatrix}
\negthinspace
\begin{pmatrix}
1&\phantom{-}1\\1&-1
\end{pmatrix}=
\tfrac{1}{2}
\begin{pmatrix}
1&2v\negthinspace-\negthinspace 1\\2v\negthinspace
-\negthinspace1&1
\end{pmatrix}
\longrightarrow\\
& \tfrac{w}{2}\begin{pmatrix} 0&1\\1&0
\end{pmatrix}
\negthinspace
\begin{pmatrix}
1&2v\negthinspace-\negthinspace 1\\2v\negthinspace
-\negthinspace1&1
\end{pmatrix}
\negthinspace
\begin{pmatrix}
0&1\\1&0
\end{pmatrix}+
\tfrac{1-w}{2}
\begin{pmatrix}
1&2v\negthinspace-\negthinspace 1\\2v\negthinspace
-\negthinspace1&1
\end{pmatrix}=\tfrac{1}{2}
\begin{pmatrix}
1&2v\negthinspace-\negthinspace 1\\2v\negthinspace
-\negthinspace1&1
\end{pmatrix}
\longrightarrow\\
&\hspace{6em}
 \tfrac{1}{4}
\begin{pmatrix}
1&\phantom{-}1\\1&-1
\end{pmatrix}
\negthinspace
\begin{pmatrix}
1&2v\negthinspace-\negthinspace 1\\2v\negthinspace
-\negthinspace1&1
\end{pmatrix}
\negthinspace
\begin{pmatrix}
1&\phantom{-}1\\1&-1
\end{pmatrix}=
\begin{pmatrix}
v&0\\
0&1-v
\end{pmatrix} .
\end{split}\vspace{1ex}
\end{equation*}
It is obvious that independently of the used tactics, human's
strategy takes the starting form. For the mixed strategy the
course of the game is described by the density operator
\begin{equation*}
\mathcal{W}=v\,\mathcal{W}_0 +(1\negthinspace-\negthinspace
v)\,\mathcal{W}_1
\end{equation*}
which also has the same diagonal form at the beginning and at the
end of the game:
\begin{equation*}
\mathcal{W}= v\, |\mathsf{0}\rangle_1\,|\mathsf{0
}\rangle_2\,{_1}\langle\mathsf{0} |\,_2\langle\mathsf{0}|+
(1\negthinspace-\negthinspace v)\,
|\mathsf{1}\rangle_1\,|\mathsf{1 }\rangle_2\,{_1}\langle\mathsf{1}
|\,_2\langle\mathsf{1}| \,.
\end{equation*}
Therefore the change of mind resulting from the female strategy
cannot lead to any additional profits. If the human using the
female tactics (that is changes his/her mind) begins the game with
the female strategy then at the end the untransparent box will be
empty and he/she will not get the content of the transparent box:
the pay-off will be minimal (0). If the human acts just the
opposite the transparent box must not be opened but nevertheless
the pay-off will be maximal (\$100000). Only if the human begins
with the female strategy and then applies the male tactics the
content  of the transparent box is accessible. If restricted to
the classical game theory Omega would have to prevent humans from
changing their minds. In the quantum domain the pay-off $M_{21}$
(female strategy and tactics) is possible (the phrase {\em la
donna mobile} gets a quantum context)\/: humans regain their free
will but they have to remember that Omega has (quantum) means to
prevent humans from profiting from altering their decisions. In
that way quantum approach allows to remove the paradox from the
rationally defined dilemma. One can also consider games with more
alternatives for the human player. The respective larger pay-off
matrices would offer even more sophisticated versions of the
Newcomb's observation. But even then there  is a quantum protocol
that guarantees that Omega keeps its promises (threats) \cite{21}.
\section{Market interpretation of the game}

It is obvious that the above scenario cannot be realized if the
actual conditions would differ from Omega's promises. For example,
Omega may not be able to predict humans intentions or its
understanding of the rules of the game differs from that implied
by their expression in human language (cultural differences).
There may be much dispute over the question {\it what Omega really
has in mind?} We would like to consider one of the variant that
may be interesting in the context of quantum market games
\cite{12,20}.
This may result from pondering over the meaning of the term {\it Omega adopts the same strategy}. \\

Players in a quantum market game sometimes buy and sometimes sell.
A demand representation of the player's strategy is a Fourier
transform of his strategy used while supplying the goods
\cite{12,11}. In a simplified model where player's strategies span
a finite dimensional Hilbert space we should apply discrete
Fourier transform which transforms the demand representation of
the strategy, being $m$-tuple of complex numbers $\langle
d|\psi\rangle$ to the supply representation given by
\begin{equation*}
\langle s|\psi\rangle=\tfrac{1}{\sqrt{m}}\sum\limits_{d=0}^{m-1}
\text{e}^{\frac{2\pi\text{i}}{m}sd} \langle d|\psi\rangle\,.
\end{equation*}

If  $m\negthinspace=\negthinspace2$ then the discrete Fourier
transform reduces to the Hadamard transform $\mathcal{F}$ which we
have already met. In our case the Hadamrd transform switches
maximally localized strategies with the the maximally indefinite
strategies and vice versa, eg\mbox{.} $\langle
d|\psi\rangle\negthinspace=\negthinspace[d\negthinspace=\negthinspace0]
\xrightarrow{\mathcal{F}}\langle
s|\psi\rangle\negthinspace=\negthinspace \tfrac{1}{\sqrt{2}}$ (the
Iverson notation \cite{22} is used: $[expression]$\/ denotes the
logical value (1 or 0) of the
sentence $expression$).\\

We introduce the nonhomogeneous complex coordinate
$z\in\overline{\mathbb{C}}$ to parameterize player's strategies
$$\mathcal{H}\ni|\psi_z\rangle:=|\mathsf{0}\rangle+z|\mathsf{1}\rangle\,.$$
If the "buying human" decides to use the strategy
$|\psi_z\rangle_1$ and the other side of the bargain (Omega) want
to play in the same way and therefore uses the supply
representation of human's strategy setting its $q$-bit to
$\mathcal{F}(|\mathsf{0}\rangle_2+z|\mathsf{1}\rangle_2)=
|\mathsf{0}\rangle_2+\tfrac{1-z}{1+z}|\mathsf{1}\rangle_2$ then
the quantum state  of the game takes the form\footnote{Cases when
the player may in fact play also "against himself" often happen in
market description: demand or supply result from self-consistent
strategy of all players. For example in a stock exchange one  big
bid or transaction can influence the whole market}
\begin{equation*}
\mathcal{W}_z=\tfrac{(1+z)(1+\overline{z})}{2(1+z\overline{z})^2}\,\bigl(
|\mathsf{0} \rangle_1+z\,|\mathsf{1}\rangle_1\bigr)
\bigl(\vphantom{i}_1\langle\mathsf{0}|+\overline{z}\,_1\langle\mathsf{1}|\bigr)
\bigl(|\mathsf{0}
\rangle_2+\tfrac{1-z}{1+z}\,|\mathsf{1}\rangle_2\bigr)
\bigl(\vphantom{i}_2\langle\mathsf{0}|+\tfrac{1-\overline{z}}{1+\overline{z}}\,_2
\langle\mathsf{1}|\bigr)\,.
\end{equation*}
Therefore, as in the previous discussion, the female strategy
gives not higher a pay-off. The expectation value of the human's
pay-off, $\text{Tr}\mathcal{M}\mathcal{W}_z$, is maximal for a
superposition of male and female strategies with phase shifted by
$\pi$  (i.e\mbox{.} for $z\negthinspace=-\negthinspace1$). In this
case the human is better off than in the previous case but she or
he must be cautious because the phase shift by $\pi$
($z\negthinspace=\negthinspace1$) does not change the respective
probabilities but result in the lowest expectation value of the
pay-off (\$500). Classical human's strategies correspond to
$z\negthinspace=\negthinspace0$ (female) and
$z\negthinspace=\pm\infty$ (male). The expectation values of the
human pay-off with respect to the adopted strategy are presented
in Figure 1.

\begin{figure}[h]
\begin{center}
\includegraphics[height=6.25cm, width=9cm]{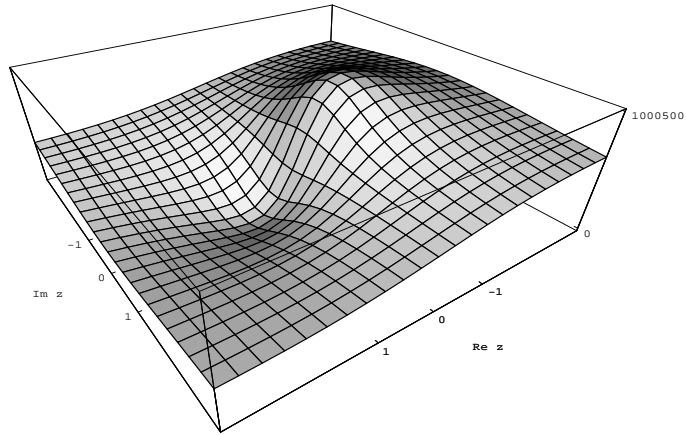}
\end{center}
\caption{The average human pay-off
$\text{Tr}\mathcal{M}\mathcal{W}_z$ in the market version of the
Newcomb game.}
 \label{hhhnnnew}
 \end{figure}
Enthusiasts for newcombmania will certainly find a lot of new
quantum solutions to the Newcomb game.

\def\urla{\href{http://econwpa.wustl.edu:8089/eps/get/papers/9904/9904004.html}{http://econwpa.wustl.edu:8089/eps/get/papers/9904/9904004.html}}
\def\urlb{\href{http://www.spbo.unibo.it/gopher/DSEC/370.pdf}{http://www.spbo.unibo.it/gopher/DSEC/370.pdf}}
\def\urlc{\href{http://www.comdig.org}{http://www.comdig.org}}

\end{document}